\documentclass[journal,twoside]{IEEEtran}
\usepackage[pdftex]{graphicx}
\setkeys{Gin}{width=0.9455\linewidth}

\usepackage[T1]{fontenc}
\usepackage{amsmath}
\usepackage[cmintegrals]{newtxmath}
\usepackage{bm}
\usepackage{nicefrac}
\usepackage[utf8]{inputenc}
\graphicspath{{./Figures/}}
\DeclareGraphicsExtensions{.pdf,.eps,.png,.jpeg}
\usepackage{color} %red, green, blue, yellow, cyan, magenta, black, white
\definecolor{mygreen}{RGB}{28,172,0} % color values Red, Green, Blue
\definecolor{mylilas}{RGB}{170,55,241}

\usepackage{slashbox}
\usepackage{breqn}
\setkeys{breqn}{labelprefix={eq:}}

\DeclareMathOperator*{\argmin}{\arg\,\min}

\newcommand{\norm}[2]{\left\lVert#1\right\rVert_{#2}}

\newcommand*{\matr}[1]{\bm{#1}}

\newcommand*{\tran}[1]{{#1}^{\mkern-1.5mu\mathsf{T}}}

\newcommand*{\herm}[1]{{#1}^{\mkern-1.5mu\mathsf{H}}}
\newcommand*{\pinv}[1]{{#1}^{\mkern-1.5mu\dagger}}

\usepackage{dblfloatfix}
\usepackage{balance}
\usepackage{cite}
\usepackage{algpseudocode}
\newcounter{tempEquationCounter} 
\newcounter{thisEquationNumber}

\usepackage{acronym}
\input{acros}

\usepackage{siunitx}
\DeclareSIUnit{\belmilliwatt}{Bm}
\DeclareSIUnit{\dBm}{\deci\belmilliwatt}
\DeclareSIQualifier\isotropic{i}
\DeclareSIQualifier\carrier{c}

\usepackage{array}

\ifCLASSOPTIONcompsoc
 \usepackage[caption=false,font=normalsize,labelfont=sf,textfont=sf]{subfig}
\else
 \usepackage[caption=false,font=footnotesize]{subfig}
\fi

\usepackage[english]{babel}

\usepackage{url}
\usepackage[nameinlink]{cleveref}
\crefname{equation}{}{}
\crefname{section}{Sec.}{Secs.}
\crefname{figure}{Fig.}{Figs.}
\usepackage{balance}
\begin{document}
% \title{\berkertitle}
\title{Reducing Precoder/Channel Mismatch and Enhancing Secrecy in Practical MIMO Systems Using Artificial Signals}

\bstctlcite{BSTcontrol}

 \author{Berker~Peköz\IEEEmembership{,~Graduate~Student~Member,~IEEE,} Mohammed~Hafez\IEEEmembership{,~Graduate~Student~Member,~IEEE,} Selçuk~Köse,~\IEEEmembership{Member,~IEEE}
 	and Hüseyin~Arslan\IEEEmembership{,~Fellow,~IEEE}%
 	\thanks{This work was supported in part by the U.S. National Science Foundation under Grant ECCS-1609581.}
 \thanks{B.~Peköz is with the Department of Electrical Engineering, University of South Florida, Tampa, FL 33620 USA (e-mail: pekoz@usf.edu).}%
 \thanks{M. Hafez is with Intel Corp., Santa Clara, CA, USA (e-mail: mhafez@mail.usf.edu).}%
 	\thanks{S.~Köse is with the Department of Electrical and Computer Engineering, University of Rochester, Rochester, NY 14627 USA (e-mail: selcuk.kose@rochester.edu).}%
 	\thanks{H.~Arslan is with the Department of Electrical Engineering, University of South Florida, Tampa, FL 33620 USA and also with the Department of Electrical and Electronics Engineering, Istanbul Medipol University, Istanbul, 34810 TURKEY (e-mail: arslan@usf.edu).}}%

\maketitle
\begin{abstract}
Practical \ac{mimo} systems depend on a predefined set of precoders to provide spatial multiplexing gain. This limitation on the flexibility of the precoders affects the overall performance. Here, we propose a transmission scheme that can reduce the effect of mismatch between users' channels and precoders. The scheme uses the channel knowledge to generate an artificial signal, which realigns the predefined precoder to the actual channel. Moreover, the scheme can provide an additional level of secrecy for the communication link. The performance of the proposed scheme is evaluated using \ac{ber}, \ac{evm}, and secrecy capacity. The results show a significant improvement for the legitimate user, along with a degradation for the eavesdropper. 
\end{abstract}
\acresetall
\IEEEpeerreviewmaketitle

% \fi
\IEEEpeerreviewmaketitle

\section{Introduction}\label{sec:intro}

\IEEEPARstart{M}{ultiple} antenna systems have been essential part of almost all current wireless systems, and will be part of any upcoming wireless standard. \Ac{mimo} systems introduce additional \acused{dof}\acp{dof} that can be utilized to provide diversity, facilitate multiplexing, or enhance secrecy.
%The capacity and available \acp{dof} of \ac{mimo} systems were thoroughly investigated, and the studies shows the significant gain that can be achieved\cite{Goldsmith03}.
Thorough investigation proves additional \acp{dof} introduced \ac{mimo} systems provide significant capacity gains\cite{Goldsmith03}.

With the expected migration towards higher frequency ranges (i.e., mmWave) in the next generation networks, another form of \ac{mimo} systems is expected to be adopted, namely, hybrid \ac{mimo}. In mmWave a larger number of antennas can be packed into smaller sizes. As promising as that sounds, that large number imposes a huge load on the system in terms of both software and hardware. Hybrid \ac{mimo} introduces a cost reduction to the system by reducing the number of used RF chains, where each subset of antennas is derived using a single RF chain \cite{Heath16}. Then, using only a set of phase-shifters, an analog beamformer is applied for that subset of antennas.

On another hand, in a fully digital or hybrid \ac{mimo}, signal precoding raises a computational complexity issue. The optimization of the precoders usually involve heavy computational processes. Moreover, the overhead to transfer the precoding information between the transmitter and receiver deems this approach unfeasible. In order to avoid both issues, the current wireless standards rely on codebooks \cite{DAHLMAN18}. The predefined codebook reduces complexity by avoiding the computational processes. Also, it reduces the overhead as the index of the used precoder is only information required to be transferred.

A downside of having a predefined codebook is the availability of such information to the public. This availability can help any malicious node in the system to receive the data correctly. This lack of information security goes against the philosophy of next generation networks. The future networks include applications with highly sensitive information (e.g., remote surgery). These applications require additional measures for information security, which brings physical-layer security to the picture. In physical layer security, the unique characteristics of the communications medium (i.e., channel) is used to protect the data from different malicious attacks (e.g., eavesdropping) \cite{Zhou13}.

To this day, the wireless community has been focusing on the design of precoders in general\cite{Dong19}, or codebooks design specifically \cite{Thoota19}. Moreover the designed codebooks usually have a single aim either enhanced achievable rate, better energy efficiency, or lower complexity \cite{Wang18}. Beside the precoding design, \ac{an} insertion approaches are used to provide some security measures \cite{Goel08}. \ac{an} approaches try to balance the trade-off between performance and security using different noise power allocation algorithms \cite{Tsai14}.

In this work, we propose two approaches to construct \acp{as}. These \acp{as} are designed to realign the codebook-based precoders to the actual \ac{mimo} communication channel. Such a design has the following benefits:
\begin{itemize}
    \item Easy direct implementation that avoids the power allocation optimization required by \ac{an} insertion algorithms.
    \item Enhanced legitimate user performance by mitigating the mismatch between codebook precoders and the actual channel.
    \item Additional layer of secrecy as the transmitted signal is constructed using the channel of the legitimate user.
    \item Keeping the same simple feedback structure conserves the adopted reduced overhead.
\end{itemize}

The rest of this paper is organized as follows: Section II provides the adopted system model. Section III introduces \ac{as} construction approaches. Gains are demonstrated in Section~IV. Finally, the paper is concluded in Section V.

Notation: Throughout this paper, vectors are represented using lowercase bold-face letters,
matrices are uppercase bold-face letters, and non-bold letters are used for scalars. The superscripts 
$\herm{\left(\cdot\right)}$, $\left(.\right)^{-1}$ stand for the conjugate-transpose, and inverse operations, respectively.
$\mathbb{C}$ represents the complex numbers domain, and $\sim\mathcal{CN}\left(\mu,\sigma^2\right)$ corresponds to complex Gaussian distributed random variable with mean $\mu$ and variance $\sigma^2$. $\norm{\cdot}{}$ corresponds to the  Euclidean norm.

%\IEEEpubidadjcol
\section{System Model}
A transmitting device, hereinafter referred to as Alice, wishes to convey information to another device, hereinafter referred to as Bob. Alice transmits the information over $N$ antennae while Bob receives the information over $M$ antennae, where $M\leq N$. The communication channel between each antenna of Alice and Bob is representable in the form of one tap over the utilized bandwidth, and is time-invariant over the transmission interval. The channel coefficient between Alice's $n$th antenna and Bob's $m$th antenna is represented in the $m$th row and $n$th column of the matrix $\matr{H}\in\mathbb{C}^{M\times N}$, and all elements are assumed to be known perfectly by Alice. The information symbols that are desired to be conveyed over a transmission interval are denoted by the vector $\matr{s}\in\mathbb{C}^{M\times 1}$.

Ideally, the \ac{mi} between the information symbols and their received counterparts is maximized if Alice precodes the symbols with $\matr{V}\in\mathbb{C}^{N\times M}$, comprising the first $M$ columns of the unitary matrix $\check{\matr{V}}\in\mathbb{C}^{N\times N}$ and Bob combines the channel outputs with the unitary matrix $\herm{\matr{U}}\in\mathbb{C}^{M\times M}$, where \cite{gallager_waveform_1968}
\begin{equation}
    \matr{H}=\matr{U}\matr{D}\herm{\check{\matr{V}}}\label{eq:channelsvd}
\end{equation}
is the \ac{svd} of the channel coefficient matrix $\matr{H}$\cite[Sec. 3]{telatar_capacity_1999}. The received symbol estimates $\hat{\matr{s}}\in\mathbb{C}^{M\times 1}$ in the ideal case are modeled as
\begin{equation}
    \hat{\matr{s}}=\matr{D}^{-1}\herm{\matr{U}}\left(\matr{H}\matr{V}\matr{s}+\matr{n}\right),\label{eq:idealtransception}
\end{equation}
where the parenthesized content is the signal received at Bob's antennae, where elements of $\matr{n}\in\mathbb{C}^{M\times 1}$ are independent and identically distributed with $\sim\mathcal{CN}\left(0,\nicefrac{1}{\gamma}\right)$ where $\gamma$ is the overall \ac{snr} of Bob for mean channel gain.

While the scheme described in \cref{eq:idealtransception} maximizes capacity and is secure, in practice, precoder-combiner matrix pair $\tilde{\matr{V}}$ and $\herm{\tilde{\matr{U}}}$ that are imperfect approximations of the similarly denoted counterparts in \cref{eq:channelsvd} may be used due to reasons made clear in \cref{sec:intro}.
Let 
\begin{equation}
	\tilde{\matr{H}}=\sqrt{\phi}\matr{H}+\left(1-\sqrt{\phi}\right)\matr{W}
\end{equation}
denote the precoder-combiner induced channel $\tilde{\matr{H}}\in\mathbb{C}^{M\times N}$ where $0\leq\phi\leq 1$ denotes the correlation between $\matr{H}$ and $\tilde{\matr{H}}$; and $\matr{W}\in\mathbb{C}^{M\times N}$ is the mismatch between them. 
Accordingly, \cref{eq:idealtransception} will hereinafter be considered as
\begin{equation}
    \hat{\matr{s}}=\matr{D}^{-1}\herm{\tilde{\matr{U}}}\left(\matr{H}\tilde{\matr{V}}\matr{s}+\matr{n}\right).\label{eq:practictransception}
\end{equation}
The $\tilde{\matr{V}}$ and $\herm{\tilde{\matr{U}}}$ matrix pairs are public knowledge, and are known by a third device with $L>N$ antennae, hereinafter referred to as Eve, that does not respect the confidentiality principle and wishes to unlawfully eavesdrop the information Alice conveys to Bob. Let $\breve{\matr{H}}\in\mathbb{C}^{L\times N}$ similarly refer to the communication channel between Alice and Eve, known perfectly by Eve, and similarly denoted and sized counterparts of elements in \cref{eq:practictransception} to other modeled properties. Eve estimates the information signals as
\begin{equation}
    \breve{\matr{s}}=\herm{\tilde{\matr{V}}}\pinv{\breve{\matr{H}}}\left(\breve{\matr{H}}\tilde{\matr{V}}\matr{s}+\breve{\matr{n}}\right),\label{eq:evesestimates}
\end{equation}
wherein $\pinv{\breve{\matr{H}}}=\tran{\breve{\matr{H}}}\left(\breve{\matr{H}}\tran{\breve{\matr{H}}}\right)^{-1}$ is the pseudo-inverse of $\breve{\matr{H}}$.

\section{Artificial Signal Construction}
Instead of transmitting the information symbols directly or precoding them with the nonideal precoder, an \ac{as} that maximizes the mutual information between $\matr{s}$ and $\hat{\matr{s}}$ can be designed. \Ac{zf} the \ac{as} such that the \ac{ls} estimates match the information symbols perfectly results in a power-unbounded \ac{as} of which power has to be downscaled to meet the transmit power requirements, hence is not an efficient way to approach the problem. The \ac{as} minimizing the instantaneous error while efficiently utilizing the transmit power can be formulated as
\begin{IEEEeqnarray}{cCLC}
\IEEEyesnumber \label{eq:precodedas}
\tilde{\matr{x}} &= & \argmin_{\matr{\xi}} & \norm{\matr{D}^{-1}\herm{\tilde{\matr{U}}}\matr{H}\tilde{\matr{V}}\matr{\xi}-\matr{s}}{}\IEEEyessubnumber\\ 
& & \text{subject to} & \norm{\matr{\xi}}{} \leq \sqrt{N}\IEEEyessubnumber
,
\end{IEEEeqnarray}
which is a convex optimization problem that can be solved computationally efficiently without introducing long processing delays\cite{gb08}.

Note that $\tilde{\matr{x}}\in\mathbb{C}^{M\times 1}$ designed in \eqref{eq:precodedas} takes the precoder into account. This is particularly useful if the precoding operation is performed in the hardware level, such as hybrid beamformers\cite{huang_hybrid_2010} or other mechanical beamformers\cite{mumcu_microfluidic_2013} such as lens array beamformers\cite{mumcu_mmwave_2018}, and these beamformers are not to be removed from the system. Hardware limitations such as the resolutions of the phase shifters, digital to analog converters (DACs) and the analog to digital converters (ADCs) at the receiver (if known) are also reflected in the precoder and combiner matrices \cite{Mendez-Rial2015}. On the other hand, if the transmitter is capable of digital beamforming, a signal design allowing the removal of the precoder from the system is possible. Accordingly,  \eqref{eq:precodedas} can be further simplified to
\begin{IEEEeqnarray}{cCLC}
\IEEEyesnumber \label{eq:noncodedas}
\matr{x} &= & \argmin_{\matr{\xi}} & \norm{\matr{D}^{-1}\herm{\tilde{\matr{U}}}\matr{H}\matr{\xi}-\matr{s}}{}\IEEEyessubnumber\\ 
& & \text{subject to} & \norm{\matr{\xi}}{} \leq \sqrt{N}\IEEEyessubnumber
,
\end{IEEEeqnarray}
which also provides more freedom as $\matr{x}\in\mathbb{C}^{N\times 1}$.

The complexity of the algorithms are not derived, but compared to the prior art, the equations involve lesser number of variables, therefore the complexity is logically expected to be lower than those already found acceptable.
\section{Results}
\begin{figure}
    \centering
    \includegraphics{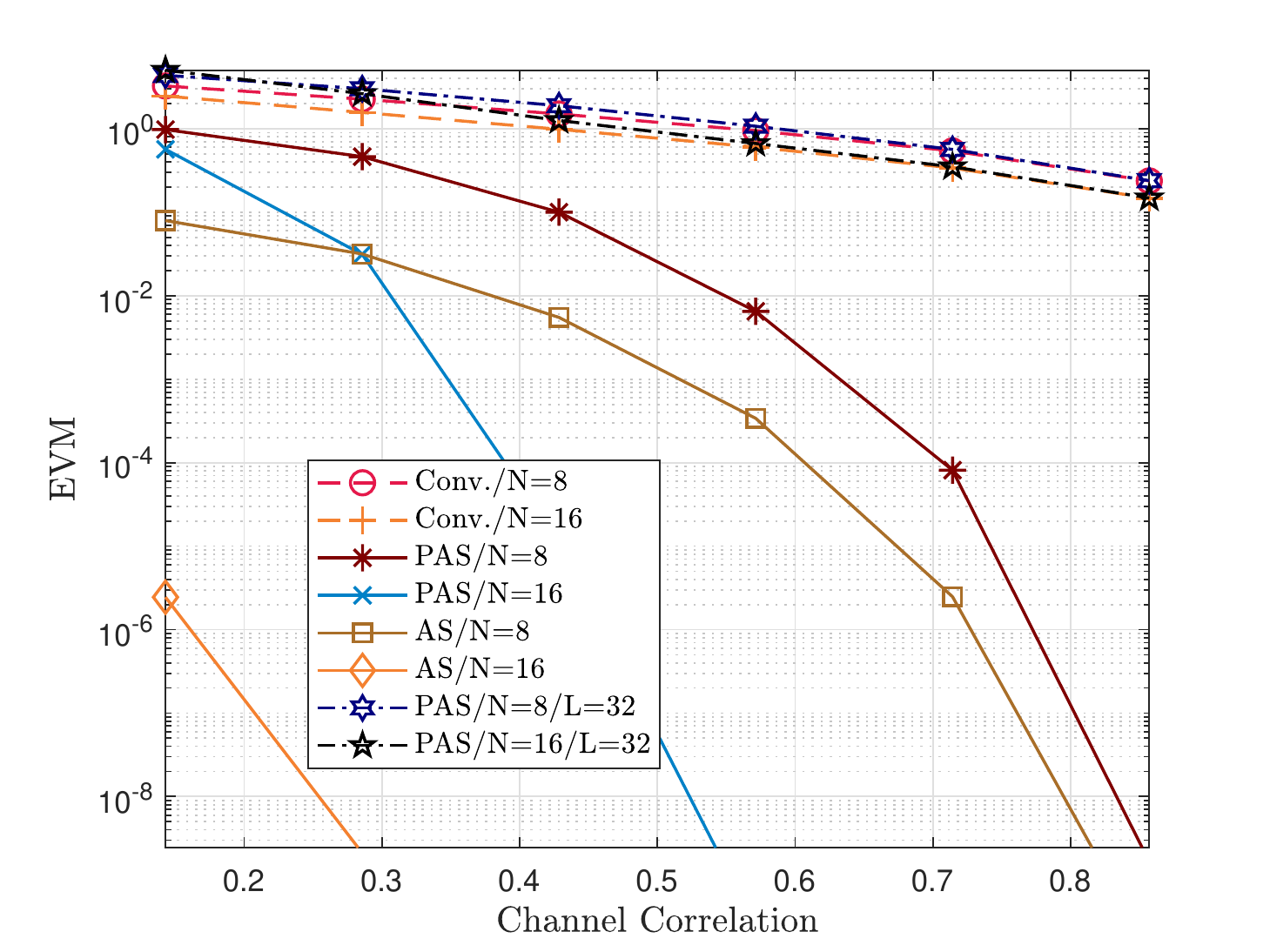}
    \caption{\Ac{evm} at various receivers in the absence of noise.}
    \label{fig:evm}
\end{figure}

\begin{figure}
    \centering
    \includegraphics{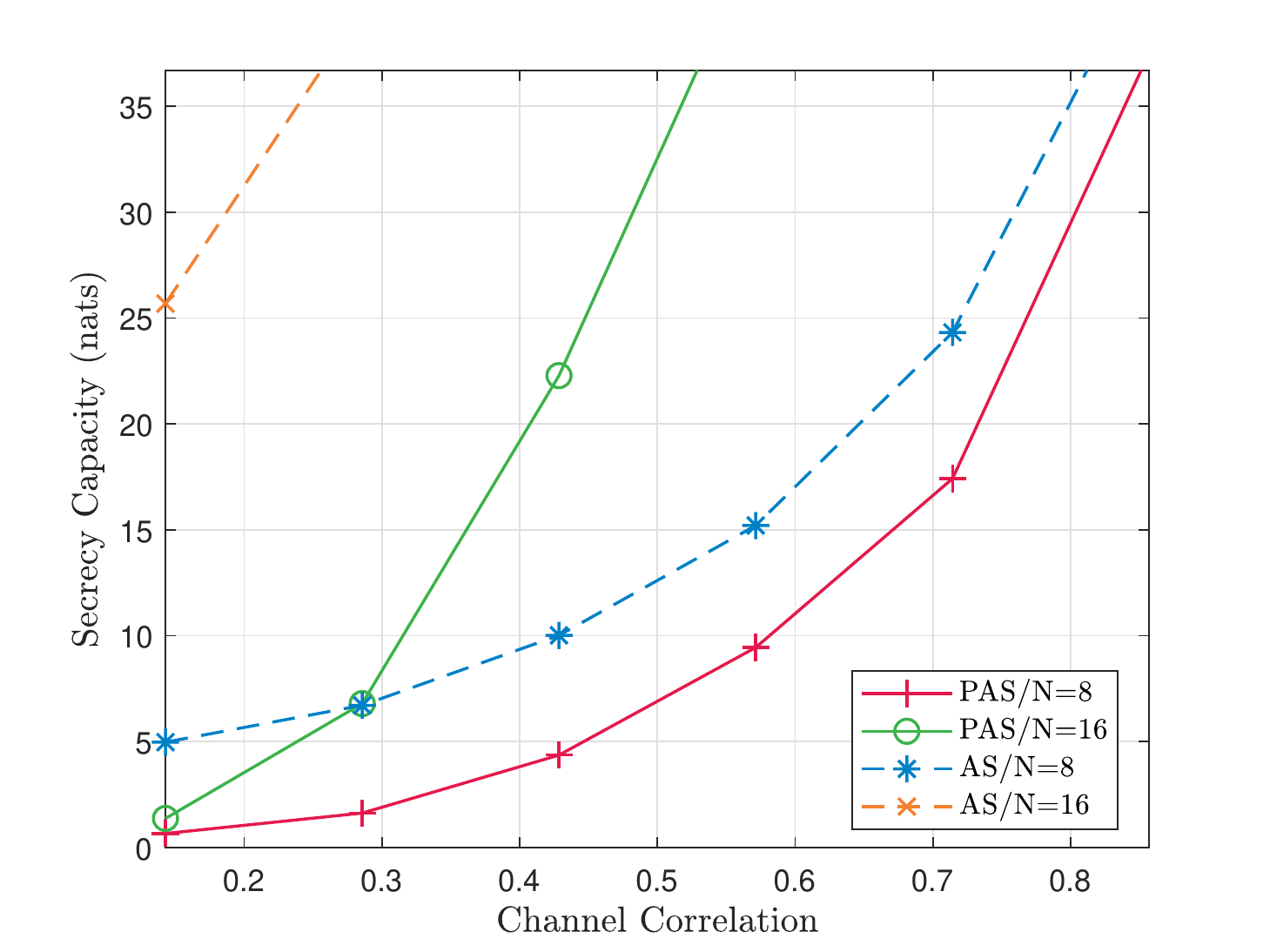}
    \caption{Secrecy capacity between Bob and Eve.}
    \label{fig:sc}
\end{figure}

The gains of the proposed technique are numerically verified by comparing the \acp{evm} and uncoded \acp{ber} at Bob and Eve as well as the secrecy capacity for conventional codebook transmission, \ac{pas} transmission per \eqref{eq:precodedas} and direct \ac{as} transmission per \eqref{eq:noncodedas} as a function of $\phi$ and $\gamma$.
%The gains provided per transmit antenna and the robustness against increasing number of eavesdropper antennas are also evaluated.
$M=4$, $N\in \left\{8,16\right\}$, $L=32$, $\matr{W}\sim\mathcal{CN}\left(0,1\right)$ and information symbols comprising $\matr{s}$ are QPSK modulated regardless of \ac{snr}, which ranges from \SIrange{0}{10}{\deci\bel}. Both \eqref{eq:precodedas} and  \eqref{eq:noncodedas} were solved using CVX, a package for specifying and solving convex programs\cite{cvx,gb08}. In the following figures, the curve is observed at Bob if only $N$ is provided, whereas it is observed at Eve if $L$ is also provided. In \cref{fig:evm,fig:sc}, $\phi=1$ (precoder perfectly aligned to the channel) results are not shown as all schemes abruptly converge as expected, which occurs at a value very different than the rest of the figure.

\cref{fig:evm} demonstrates the change in the precoding-combining quality provided by the proposed technique by comparing the \ac{evm} at various receivers in the absence of noise as a function of $\phi$. It is seen that the conventional scheme does not yield a waterfall gain unless $\phi\to 1$, and doubling the number of transmitter antennae reduces EVM by about \SI{1}{\deci\bel}. The proposed algorithms, however, present waterfalling EVM schemes at the legitimate receiver for any channel correlation and greatly outperform the conventional scheme for any nonunitary $\phi$. While the number of transmitter antennae is the most significant factor in reducing EVM for both proposed algorithms, AS significantly outperforms PAS due to the increased level of flexibility for lower $\phi$ values while the difference narrows as $\phi$ increases. In the meantime, both proposed algorithms (AS not drawn due to overlapping) limit the EVM performance at Eve to that provided by the conventional schemes at the legitimate receiver for $\phi\neq 1$. The EVM at Eve for conventional transmission is mathematically insignificant for all investigated valid number of antennae combinations, hence is not shown.

\cref{fig:sc} shows the secrecy capacity comprising the difference of capacities between Bob and Eve as a function of $\phi$ in the absence of noise. The findings of \cref{fig:evm} are confirmed, AS is more secure than PAS at lower $\phi$ as a result of the additional flexibility, which is later dominated by $N$ as $\phi$ increases due to additional diversity. The secrecy capacity increases up to $\phi\to 85\%$ and diminishes to zero beyond higher correlations as Eve's capacity increases. The secrecy capacity of conventional transmission is zero hence is not shown.

\begin{figure}
    \centering
    \includegraphics{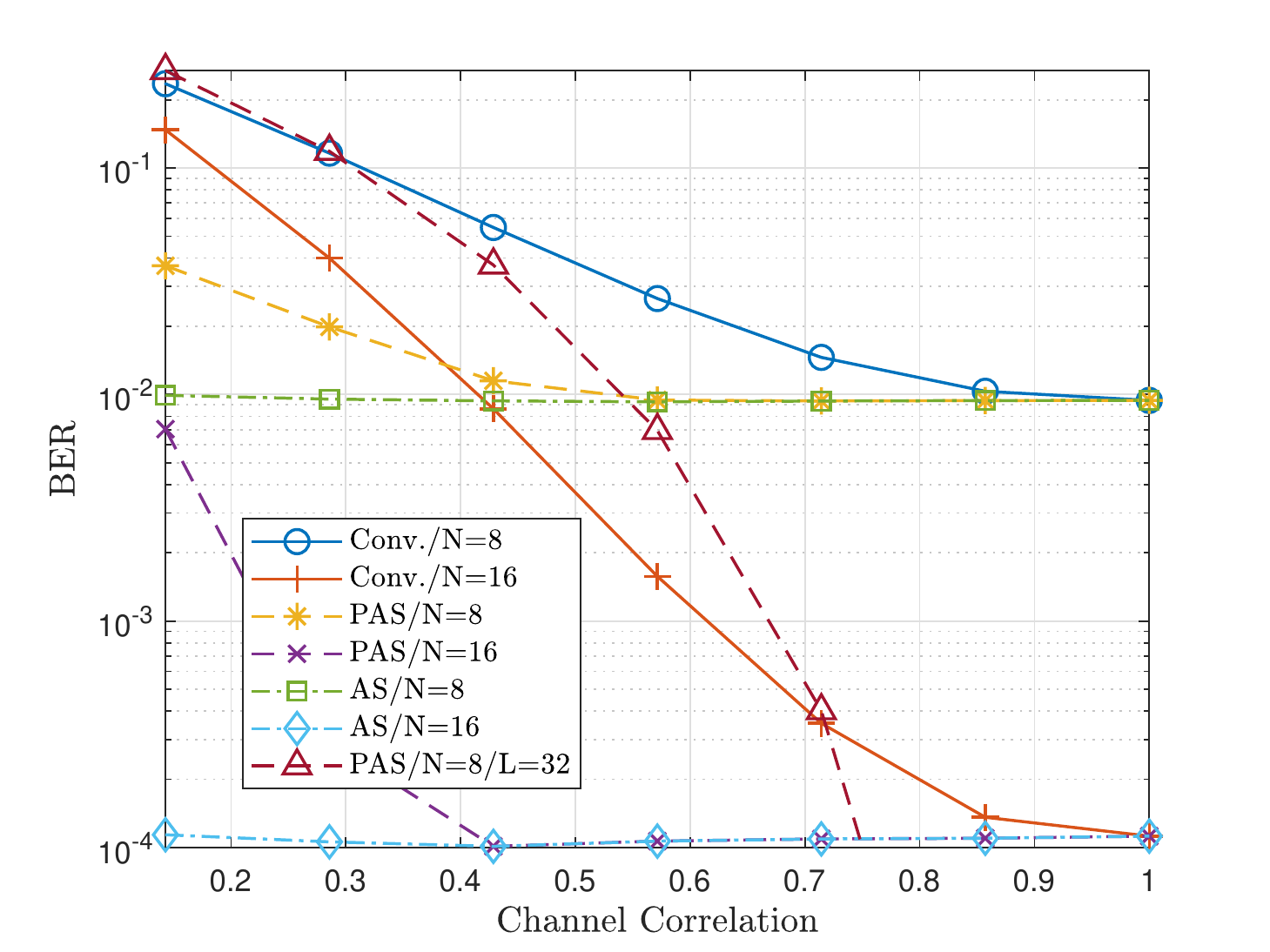}
    \caption{\Ac{ber} at various receivers for \ac{snr}=\SI{3}{\deci\bel}.}
    \label{fig:bercorr}
\end{figure}

\cref{fig:bercorr} shows the BER as a function of $\phi$ for \SI{3}{\deci\bel} SNR. Under noisy reception, the performance of AS becomes independent of $\phi$ as the introduced flexibility allows matching the exact channel at any $\phi$ and is dominated by the diversity provided by the number of antennas. The performance of PAS converges to that of AS, and the convergence $\phi$ value decreases with increasing the number of antennas as the increased diversity allows easier matching. The performance of conventional transmission converges to that of proposed schemes as $\phi\to 1$, and the proposed schemes have significant advantage otherwise. The performance at Eve waterfalls with $\phi$, confirming that $\phi$ is the dominating factor in noisy reception as the high diversity greatly improves SNR.
%It is visible that $N$ plays the most important role in all cases, however, even the PAS scheme at least halves the BER of conventional transmission using half the number of transmitter antennae, whereas the gain of AS is of greater magnitude. Gain of PAS diminishes with increasing $\phi$ (before waterfalling abruptly at $\phi=1$) whereas the performance of AS continues growing linearly in dB scale without bottlenecking. The performance at Eve remains flat at finite SNR for both PAS and AS (overlaps, removed for clarity), as similarly shown in \cref{fig:evm}, remaining close to the performance of Bob for conventional transmission.

\begin{figure}
    \centering
    \includegraphics{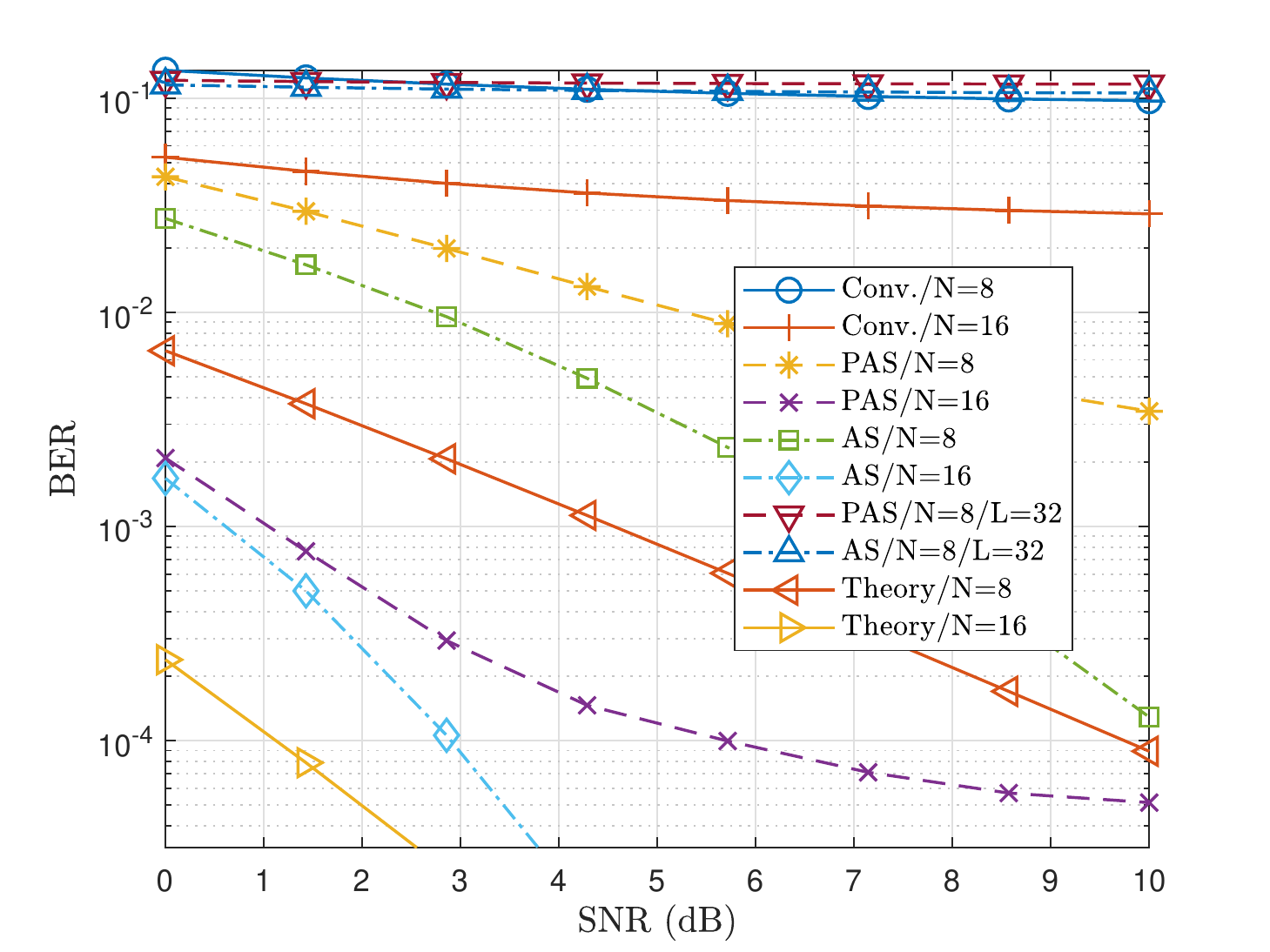}
    \caption{\Ac{ber} at various receivers for $\phi=30\%$.}
    \label{fig:ber28}
\end{figure}

\begin{figure}
    \centering
    \includegraphics{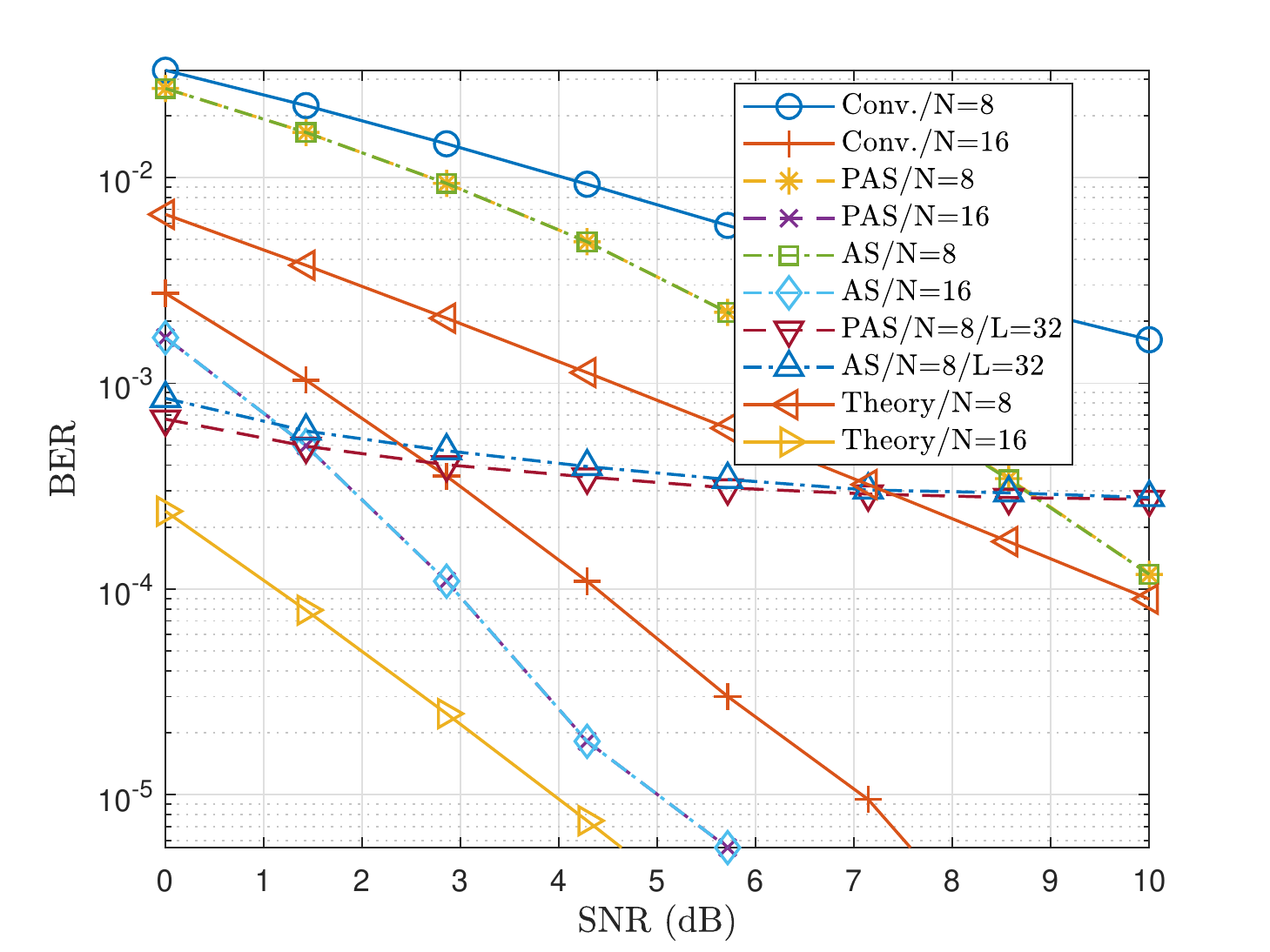}
    \caption{\Ac{ber} at various receivers for $\phi=70\%$.}
    \label{fig:ber71}
\end{figure}

\cref{fig:ber28,fig:ber71} show the BER as a function of SNR for $\phi=30\%$ and $\phi=70\%$, respectively. The theoretical limits derived in \cite{telatar_capacity_1999} for the optimally precoded and combined transmission described by \cref{eq:idealtransception} are also presented for comparison. The performance of the conventional scheme is bottlenecked by $\phi$ independent of SNR for low $\phi$, whereas the performance of PAS increases before being bottlenecked by $\phi$ for low $\phi$ values. On the other hand, the performance of AS converges to the theoretical limit with increasing SNR regardless of $\phi$, a phenomenon commonly observed in fading channels with suboptimum equalization, of which optimization falls beyond the scope of this article.
The gap between theoretical limit and AS performance is independent of $\phi$ for $\phi>30\%$ in accord with \cref{fig:bercorr}. Furthermore, the gap between PAS and AS closes as $\phi$ increases in accord with \cref{fig:bercorr}. The BER at Eve, which has 8 times the diversity of Bob, remains bottlenecked by $\phi$ and does not depend much on SNR for both proposed schemes at both $\phi$ values, showing that the security gap between the two proposed signal designs is insignificant in practical SNRs.
%It is seen that the BER is independent of SNR for the AS scheme, but depends on the number of transmitter antennae as expected (BER for AS scheme with $N=16$ is similarly constant at $\sim 9\times 10^{-5}$, which is not shown for clarity). The PAS scheme converges to the same BER as AS at \SI{6}{\deci\bel}, which is achieved at \SI{10}{\deci\bel} if the conventional codebook transmitter is used. Although \cref{fig:evm,fig:sc} suggest Eve's performance depends on the scheme at infinite SNR, the BER at Eve is independent of whether PAS or AS is used in the practically realistic SNR values. The BER performance of Eve overtakes Bob at higher SNRs as Eve has 8 times the antennas as Bob, increasing the diversity gain in finite SNR.
%\cref{fig:ber71} shows the BER as a function of SNR for $\phi=70\%$. Jointly considering \cref{fig:ber28,fig:ber71} suggests that the SNR values at which the BER performance of conventional scheme, PAS and AS converge is independent of $\phi$, while the gap between AS over others enlarges as $\phi$ increases. The gap between PAS and eavesdropper also widens with increasing $\phi$.

\balance
% \vspace{-1em}
\section{Concluding Remarks}
We proposed two different approaches to construct an artificial signal, which can mitigate the mismatch between the channel and the codebook-based precoders. The constructed signal is able to reduce the \ac{ber} experienced by legitimate user, while keeping eavesdropper's \ac{ber} at a high level. The secrecy performance of PAS and AS are theoretically different at infinite SNR, but for practical SNR values the approaches are indifferent. If the hardware allows full digital beamforming this increases the capacity at the intended receiver, whereas the eavesdropper will keep believing a precoder is used, creating additional confusion. The performance of precoded method converges to that of nonprecoded beyond a certain correlation, of which value decreases as diversity rank increases. Even though the proposed algorithms still enhances the performance in case of low or no correlation, we suggest  that it would be more beneficial to keep the operating point around $0.6$ to $0.9$. At that operating range, both the performance of legitimate user and the secrecy gap experience a satisfying enhancement.

\ifCLASSOPTIONcaptionsoff
  \newpage
\fi



%\bibliographystyle{IEEEtran}

\bibliographystyle{IEEEtran}
\bibliography{IEEEabrv,berker,3gpp_38-series}

\begin{thebibliography}{10}
\providecommand{\url}[1]{#1}
\csname url@samestyle\endcsname
\providecommand{\newblock}{\relax}
\providecommand{\bibinfo}[2]{#2}
\providecommand{\BIBentrySTDinterwordspacing}{\spaceskip=0pt\relax}
\providecommand{\BIBentryALTinterwordstretchfactor}{4}
\providecommand{\BIBentryALTinterwordspacing}{\spaceskip=\fontdimen2\font plus
\BIBentryALTinterwordstretchfactor\fontdimen3\font minus
  \fontdimen4\font\relax}
\providecommand{\BIBforeignlanguage}[2]{{%
\expandafter\ifx\csname l@#1\endcsname\relax
\typeout{** WARNING: IEEEtran.bst: No hyphenation pattern has been}%
\typeout{** loaded for the language `#1'. Using the pattern for}%
\typeout{** the default language instead.}%
\else
\language=\csname l@#1\endcsname
\fi
#2}}
\providecommand{\BIBdecl}{\relax}
\BIBdecl

\bibitem{Goldsmith03}
A.~{Goldsmith}, S.~A. {Jafar}, N.~{Jindal}, and S.~{Vishwanath}, ``Capacity
  limits of {MIMO} channels,'' \emph{{IEEE} J. Sel. Areas Commun.}, vol.~21,
  no.~5, pp. 684--702, June 2003.

\bibitem{Heath16}
R.~W. {Heath}, N.~{González-Prelcic}, S.~{Rangan}, W.~{Roh}, and A.~M.
  {Sayeed}, ``An overview of signal processing techniques for millimeter wave
  {MIMO} systems,'' \emph{{IEEE} J. Sel. Topics Signal Process.}, vol.~10,
  no.~3, pp. 436--453, April 2016.

\bibitem{DAHLMAN18}
E.~Dahlman, S.~Parkvall, and J.~Sköld, ``Chapter 11 - multi-antenna
  transmission,'' in \emph{5G NR: the Next Generation Wireless Access
  Technology}, E.~Dahlman, S.~Parkvall, and J.~Sköld, Eds.\hskip 1em plus
  0.5em minus 0.4em\relax Academic Press, 2018, pp. 225 -- 240.

\bibitem{Zhou13}
X.~Zhou, L.~Song, and Y.~Zhang, \emph{Physical layer security in wireless
  communications}.\hskip 1em plus 0.5em minus 0.4em\relax Boca Raton, FL: CRC
  Press, 2013.

\bibitem{Dong19}
F.~{Dong}, W.~{Wang}, and Z.~{Wei}, ``Low-complexity hybrid precoding for
  multi-user {MmWave} systems with low-resolution phase shifters,''
  \emph{{IEEE} Trans. Veh. Technol.}, vol.~68, no.~10, pp. 9774--9784, Oct
  2019.

\bibitem{Thoota19}
S.~S. {Thoota}, P.~{Babu}, and C.~R. {Murthy}, ``Codebook based precoding and
  power allocation for {MU-MIMO} systems for sum rate maximization,''
  \emph{{IEEE} Trans. Commun.}, pp. 1--1, 2019.

\bibitem{Wang18}
X.~{Wang}, Y.~{Wang}, W.~{Ni}, R.~{Sun}, and S.~{Meng}, ``Sum rate analysis and
  power allocation for massive {MIMO} systems with mismatch channel,''
  \emph{{IEEE} Access}, vol.~6, pp. 16\,997--17\,009, 2018.

\bibitem{Goel08}
S.~{Goel} and R.~{Negi}, ``Guaranteeing secrecy using artificial noise,''
  \emph{{IEEE} Trans. Wireless Commun.}, vol.~7, no.~6, pp. 2180--2189, June
  2008.

\bibitem{Tsai14}
S.~{Tsai} and H.~V. {Poor}, ``Power allocation for artificial-noise secure
  {MIMO} precoding systems,'' \emph{{IEEE} Trans. Signal Process.}, vol.~62,
  no.~13, pp. 3479--3493, July 2014.

\bibitem{gallager_waveform_1968}
R.~G. Gallager, ``\BIBforeignlanguage{English}{Waveform {Channels}},'' in
  \emph{\BIBforeignlanguage{English}{Information {Theory} and {Reliable}
  {Communication}}}.\hskip 1em plus 0.5em minus 0.4em\relax New York: Wiley,
  1968, pp. 355--441.

\bibitem{telatar_capacity_1999}
E.~Telatar, ``Capacity of multi-antenna gaussian channels,'' \emph{Eur. Trans.
  Telecommun.}, vol.~10, no.~6, pp. 585--595, 1999.

\bibitem{gb08}
M.~Grant and S.~Boyd, ``Graph implementations for nonsmooth convex programs,''
  in \emph{Recent Advances in Learning and Control}, ser. Lecture Notes in
  Control and Information Sciences, V.~Blondel, S.~Boyd, and H.~Kimura,
  Eds.\hskip 1em plus 0.5em minus 0.4em\relax Springer-Verlag Limited, 2008,
  pp. 95--110.

\bibitem{huang_hybrid_2010}
X.~{Huang}, Y.~J. {Guo}, and J.~D. {Bunton}, ``A hybrid adaptive antenna
  array,'' \emph{{IEEE} Trans. Wireless Commun.}, vol.~9, no.~5, pp.
  1770--1779, May 2010.

\bibitem{mumcu_microfluidic_2013}
A.~A. {Gheethan}, M.~C. {Jo}, R.~{Guldiken}, and G.~{Mumcu}, ``Microfluidic
  based {Ka}-band beam-scanning focal plane array,'' \emph{{IEEE} Antennas
  Wireless Propag. Lett.}, vol.~12, pp. 1638--1641, 2013.

\bibitem{mumcu_mmwave_2018}
G.~{Mumcu}, M.~{Kacar}, and J.~{Mendoza}, ``Mm-{Wave} beam steering antenna
  with reduced hardware complexity using lens antenna subarrays,'' \emph{{IEEE}
  Antennas Wireless Propag. Lett.}, vol.~17, no.~9, pp. 1603--1607, Sep. 2018.

\bibitem{Mendez-Rial2015}
R.~Méndez-Rial, C.~Rusu, N.~González-Prelcic, and R.~W. Heath,
  ``Dictionary-free hybrid precoders and combiners for {mmWave} {MIMO}
  systems,'' in \emph{Proc. 2015 {IEEE} 16th {Int}. {Workshop} {Signal}
  {Process}. {Advances} in {Wirel.} {Commun}.}, Stockholm, SE, Jun. 2015, pp.
  151--155.

\bibitem{cvx}
M.~Grant and S.~Boyd, ``{CVX}: {MATLAB} software for disciplined convex
  programming, version 2.1,'' \url{http://cvxr.com/cvx}, Mar. 2014.

\end{thebibliography}

\end{document}